%% file: ojits-template.tex
\documentclass{IEEEoj}
\usepackage{cite}
\usepackage{amsmath,amssymb,amsfonts}
\usepackage{algorithmic}
\usepackage{graphicx,color}
\usepackage{orcidlink}
\usepackage[nolist]{acronym}
\input{acronyms}
\usepackage{bm}
\usepackage{booktabs}
\usepackage{gensymb}
\usepackage{tabularx}
\usepackage{multirow}
\usepackage{textcomp}
\def\BibTeX{{\rm B\kern-.05em{\sc i\kern-.025em b}\kern-.08em
    T\kern-.1667em\lower.7ex\hbox{E}\kern-.125emX}}
\AtBeginDocument{\definecolor{ojcolor}{cmyk}{0.93,0.59,0.15,0.02}}
\def\OJlogo{\vspace{-14pt}\includegraphics[height=28pt]{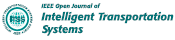}}
\begin{document}
\receiveddate{XX Month, XXXX}
\reviseddate{XX Month, XXXX}
\accepteddate{XX Month, XXXX}
\publisheddate{XX Month, XXXX}
\currentdate{XX Month, XXXX}
\doiinfo{OJITS.2022.1234567}

\title{Scalable Intention Sharing for ETSI VAMs}

\makeatletter
\renewcommand{\authorrefmark}[1]{\textsuperscript{#1}}
\makeatother

\author{\uppercase{Felipe E. Valle}\orcidlink{0000-0002-2995-8322}\authorrefmark{1},
\uppercase{Oscar Amador}\orcidlink{0000-0003-0887-5226}\authorrefmark{1}, \uppercase{Johan Thunberg}\orcidlink{0000-0002-9738-4148}\authorrefmark{2} \IEEEmembership{Member, IEEE}, \uppercase{Elena Haller}\orcidlink{0009-0004-1551-3201}\authorrefmark{1}, and \uppercase{Alexey Vinel}\orcidlink{0000-0003-4894-4134}\authorrefmark{3}
\IEEEmembership{Senior Member, IEEE}}
\affil{School of Information Technology, Halmstad University, SE-301 18 Halmstad}
\affil{Department of Electrical and Information Technology (EIT), Lund University, SE-22100 Lund}
\affil{Karlsruhe Institute of Technology, 76131 Karlsruhe, Germany}
\corresp{CORRESPONDING AUTHOR: Felipe E. Valle (e-mail: felipe.valle@hh.se).}
\authornote{``This work was supported in part by the ELLIIT network through project 6G-V2X "6G Wireless, sub-project: Vehicular Communications", and the Swedish Transport Administration through the project \textit{''Here I go'' – avancerade funktioner för VRU-medvetenhetsprotokoll i C-ITS} (TF 2023/104170).''}
\markboth{Scalable Intention Sharing for ETSI VAM}{Author \textit{Valle et al.}}

\begin{abstract}
Efficient maneuver coordination in dense V2X environments requires accurate short-term prediction while maintaining low communication and computational overhead. Current \ac{ETSI}-compliant approaches rely on intention detection and trajectory vector transmission, which scale poorly with neighborhood size and prediction horizon. This paper revisits maneuver coordination from an intention sharing perspective and investigates geometric encodings that enable scalable communication. First, we analyze three \ac{ETSI}-compliant encodings, trajectory vectors, $N$-polygons, and uncertainty ellipses, through complexity analysis and simulation-based CPU measurements. Results show that uncertainty ellipses reduce computational complexity by an order of magnitude compared with trajectory vectors while maintaining a constant message size. Building on this, an \acl{EKF} is used to generate short-horizon predictions, which are encoded as uncertainty ellipses to represent the intended maneuver.  The prediction pipeline is evaluated using real-world GNSS trajectories collected from cyclist maneuvers on a controlled test track, demonstrating that the approach achieves reliable multi-second prediction horizons while maintaining scalability for dense V2X environments.
\end{abstract}

\begin{IEEEkeywords}
Cooperative Intelligent Transport Systems (C-ITS), Connected Vehicles, Vehicular Communications (V2X), Road Safety, \acp{VRU}.
\end{IEEEkeywords}


\maketitle

\section{Introduction}
\label{sec:introduction}

\PARstart{E}{fficient} maneuver coordination in cooperative \ac{V2X} systems requires reliable short-term prediction of road-user motion while maintaining low communication and computational overhead. Through the exchange of information about their current state and intended actions, connected road users can anticipate future behavior and support safer cooperative interactions in dynamic traffic environments.

Two general approaches exist for this purpose: intention detection and intention sharing. In intention detection, each agent independently predicts the future trajectories of nearby users from observed motion histories, which can lead to redundant computation and scalability challenges in dense traffic scenarios. In contrast, intention sharing allows each road user to predict its own short-term maneuver and broadcast this information to neighboring agents.

In Europe, intention sharing is supported by the \acf{ETSI} \ac{VBS} through \acp{VAM}, which allow predicted trajectories to be transmitted as coordinate sequences. However, transmitting these sequences introduces communication overhead that grows with the prediction horizon. To mitigate this issue, the \ac{ETSI} standard restricts the size and frequency of the predicted path containers, limiting their effectiveness for highly dynamic maneuver coordination.


In our previous work~\cite{valle:hal-05424230}, we studied how message size variations affect the quality and quantity of information used to convey intention. Based on those results, we proposed sending trajectories in every \ac{VAM} to avoid size fluctuations that may create access-layer issues, in line with~\cite{Anupama2022}. We also compared several geometric representations for encoding predicted trajectories or geographic areas, trajectory vectors, uncertainty ellipses, and $N$-polygons, and found that area-based representations exhibit favorable theoretical scalability properties. However, that analysis relied on polynomial least-squares prediction, which can exhibit poor extrapolation behavior when modeling highly dynamic motion patterns such as cyclist maneuvers \cite{Hastie2009ESL}.

Building upon that foundation, this paper introduces three extensions. First, we complement asymptotic complexity analysis with simulation-based CPU operation measurements to quantify the practical computational burden of different trajectory representations under varying traffic densities and prediction horizons. Second, we replace polynomial fitting with an \ac{EKF} predictor, which provides a lightweight model-based framework for short-horizon motion estimation suitable for embedded V2X devices. Third, we experimentally evaluate the prediction pipeline using real-world cyclist trajectories collected during controlled test-track experiments.

The main contributions of this work are:
\begin{itemize}
\item A comparative evaluation of ETSI-compliant encodings—trajectory vectors, uncertainty ellipses, and $N$-polygons—with respect to their computational complexity and communication scalability.
\item An \ac{EKF}-based short-horizon prediction pipeline enabling real-time generation of motion samples for maneuver encoding on resource-constrained \ac{VRU} devices.
\item An experimental assessment using real-world cyclist trajectories, including highly non-stationary maneuvers, to evaluate prediction robustness and practical intention-sharing horizons.
\end{itemize}

The remainder of this paper is organized as follows. Section \ref{sec:related_work} reviews related work. Section \ref{sec:model_scenario} presents the cooperative V2X system model and introduces the intention sharing framework. Section \ref{sec:framework} describes the \ac{ETSI} \ac{VAM} architecture. Section \ref{sec:prediction}  introduces the \ac{EKF}-based trajectory prediction method. Section \ref{sec:representation} discusses the uncertainty ellipse encoding and trajectory representations. Section \ref{sec:evaluation} evaluates the framework using simulations and real-world data. Section \ref{sec:conclusion} concludes the paper.

\section{Related Work}
\label{sec:related_work}
Maneuver coordination in cooperative intelligent transportation systems has traditionally been addressed through intention detection mechanisms, as reflected in the ETSI specifications for \acp{VAM}\cite{etsiVAM}. In this context, substantial research has focused on inferring future actions from observed motion patterns. Schulz \emph{et al.}\cite{schulz2015pedestrian} proposed a latent-dynamic conditional random field framework for pedestrian intention recognition in urban crossing scenarios and later extended it with a multiple-model filtering scheme that jointly performs intention recognition and trajectory prediction\cite{schulz2015controlled}. Similarly, Quintero \emph{et al.}\cite{quintero2015pedestrian} combined motion dynamics with behavioral classification to improve pedestrian path prediction. For cyclists, Pool \emph{et al.}\cite{pool2017road} incorporated road topology information into route prediction models, while Benterki \emph{et al.}\cite{benterki2019lane} applied machine learning techniques to detect lane-change intentions in motor vehicles. Collectively, these approaches aim to infer the internal state or behavioral class of traffic participants from historical observations.

Despite their effectiveness, intention detection schemes inherently depend on transmitting motion histories, such as the sequence of recent trajectory points defined in the \ac{ETSI} standard\cite{etsiVAM}. In dense traffic scenarios, the inclusion of past and predicted path containers increases communication overhead and leads to variable packet sizes. Such variability adversely affects medium access performance: in IEEE 802.11-based systems it increases collision probability, while in cellular V2X it complicates radio resource allocation and scheduling\cite{Anupama2022}. Thus, substantial effort has been devoted to optimizing \ac{VAM} triggering conditions and message generation policies to reduce channel load while preserving situational awareness\cite{martinperez2023, silas2024}. Uniform and compact message structures have been shown to improve key performance indicators such as \ac{PDR} and \ac{AoI} \cite{tesis_johan}. However, \acp{VAM} carrying trajectory containers exhibit inherent payload size variability, which prevents such uniform behavior.


Furthermore, the joint design of motion prediction and trajectory encoding for scalable intention sharing has received limited attention. Existing work typically focuses either on trajectory prediction accuracy or on message dissemination strategies, leaving the interaction between prediction algorithms and geometric motion representations largely unexplored. This work addresses this gap by combining short-horizon EKF prediction with uncertainty ellipse encoding for efficient maneuver intention sharing using \acp{VAM}.

\section{System Model and Scenario}
\label{sec:model_scenario}

\begin{figure}
    \centering
    \includegraphics[width=\linewidth]{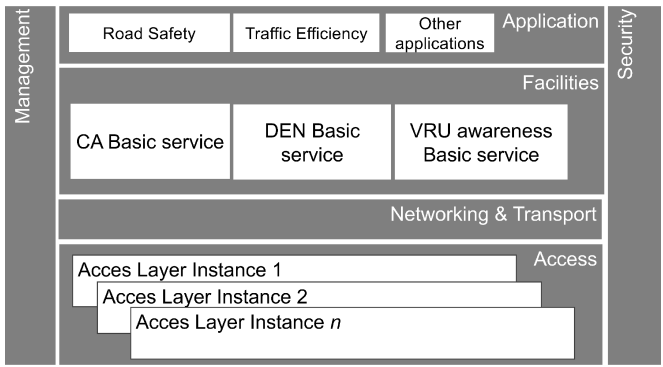}
    \caption{ETSI C-ITS Architecture}
    \label{fig:simple_arch}
\end{figure}

This section describes the cooperative \ac{V2X} scenario considered in this work. We outline the cooperative awareness setting involving vehicles and \acp{VRU}, and discuss the two main approaches for anticipating future motion: intention detection and intention sharing.


\subsection{Cooperative V2X Scenario}

We consider the \ac{ETSI} \ac{C-ITS} environment, illustrated in Fig. \ref{fig:simple_arch}. 
In particular, the focus of this work is on \acp{EPAC}, such as e-bikes and e-scooters, equipped with lightweight on-board communication units capable of broadcasting \acp{VAM}.

Each road user periodically broadcasts its kinematic state, including position, speed, and heading, allowing neighboring agents to maintain situational awareness. The communication range typically encompasses multiple nearby road users, forming a dynamic neighborhood whose size depends on traffic density and wireless channel conditions.

Within this cooperative environment, each participant must anticipate the future behavior of surrounding road users to safely perform the \ac{DDT}, especially in scenarios involving vulnerable road users whose trajectories may change rapidly.


\subsection{Passive and Active VRU Awareness}

Two complementary mechanisms allow vehicles to become aware of \acp{VRU} during the execution of the \ac{DDT}.

The first mechanism relies on on-board sensing and perception systems. In this passive approach, the vehicle detects nearby road users through sensors, as part of the \ac{OEDR} task. The vehicle must then infer the future motion of detected objects based on their observed trajectories.

The second mechanism relies on communication between connected agents. In this active approach, road users directly broadcast their position and motion information through V2X messages. When \acp{VRU} themselves are connected, for instance through embedded communication units on bicycles or scooters, they can actively transmit their kinematic state and potentially their predicted motion.

While passive sensing is essential for detecting unconnected road users, active communication can significantly extend the awareness horizon and provide additional information about motion intent. Consequently, combining perception with cooperative communication is considered a key enabler of future \ac{CCAM} systems.

\subsection{Intention Detection vs. Intention Sharing}

Once neighboring road users have been detected via sensor or through communication, their future behavior must be estimated in order to support safe maneuvering.

The older and more studied approach used to achieve this is \emph{intention detection}, in which each agent independently predicts the future trajectories of surrounding road users based on their observed motion histories. In this paradigm, an ego road user collects recent trajectory samples from its neighbors and applies prediction algorithms to estimate their future motion.

An alternative approach is \emph{intention sharing}. In this paradigm, each road user predicts its own short-term future trajectory and broadcasts that information to its neighbors. Instead of reconstructing future motion from past observations, receiving agents directly obtain the predicted maneuver from the originating road user.

Intention sharing has several conceptual advantages. Each agent possesses the most accurate knowledge of its own motion state, route-following strategy, and local control actions. Consequently, self-prediction leads to more reliable short-term trajectory estimates while avoiding redundant computation across the network. Furthermore, intention sharing aligns naturally with cooperative traffic interaction models, where agents exchange information about their intended actions in order to coordinate maneuvers such as merging, lane changes, or intersection crossing.

\subsection{Computational Implications in Dense Traffic}

The choice between intention detection and intention sharing has important implications for computational scalability. Consider a scenario with $m$ ego agents, each interacting with $n$ neighboring road users. Under the intention detection paradigm, every agent must estimate the future motion of all neighbors independently. The total computational effort for prediction at the system level therefore scales as $\mathcal{O}(mn)$.

In contrast, under the intention sharing paradigm each road user is responsible only for predicting its own trajectory. The predicted maneuver is then broadcast to neighbors, who simply consume the transmitted information. In this case, the prediction effort scales as $\mathcal{O}(n)$ since only one prediction per road user is required.

In dense traffic environments, where the number of neighboring agents can be large, this reduction in redundant computation can significantly improve scalability. However, efficient intention sharing also requires compact representations of predicted motion that can be transmitted within the constraints of V2X communication protocols.

In the following section, we describe how predicted motion can be transmitted using the \ac{ETSI} \ac{VBS} and discuss the associated message structures and communication constraints.

\section{ETSI VAM Intention Sharing Framework}
\label{sec:framework}

\label{subsubsec:vam_prediction}
\begin{figure}
    \centering
    \includegraphics[width=\linewidth]{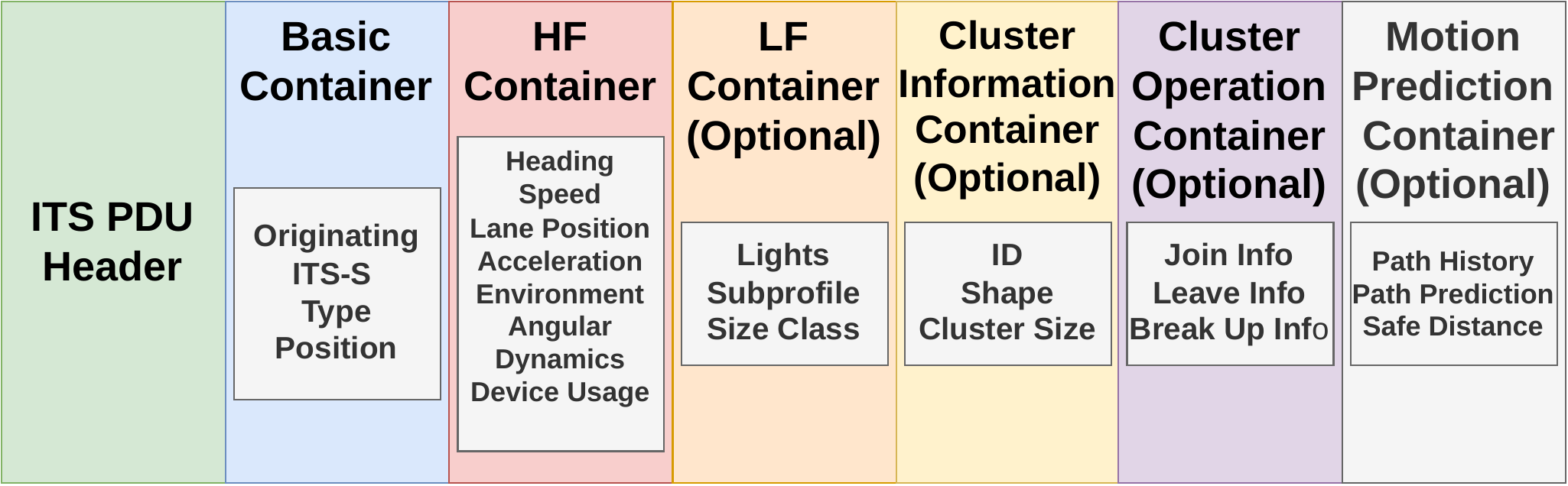}
    \caption{VAM structure}
    \label{fig:vam_structure}
\end{figure}

Within the \ac{C-ITS} framework introduced in Section \ref{sec:model_scenario} \acf{VRU} on-board devices implement the \ac{ETSI} \ac{VBS}\cite{etsi-va} to periodically broadcast \acp{VAM} containing their current kinematic state and optional motion prediction information.

The optional path prediction container, shown in Fig.~\ref{fig:vam_structure}, enables a \ac{VRU} on-board device to share information about its expected future trajectory. In the standard representation, this information is typically expressed as a sequence of predicted trajectory points (up to $15$ plus confidence intervals). While this representation provides detailed information about the predicted motion, the number of transmitted trajectory points varies with the prediction horizon and the complexity of the motion, increasing message size and processing overhead.

Previous works \cite{Anupama2022,tesis_johan} show that this size variability may lead to inefficient use of communication resources in dense traffic environments, particularly when multiple \acp{VRU} simultaneously transmit prediction information. This issue becomes even more pronounced when intention detection is required, since \acp{VRU} would also need to include the optional path history container, which may contain up to $23$ previously observed trajectory points.

To address these challenges, the proposed framework enables \acp{VRU} to share motion intention information using a compact geometric representation with fixed message size. Instead of transmitting a variable number of trajectory points, the predicted motion is encapsulated in an uncertainty ellipse describing the expected future region occupied by the cyclist. The ellipse parameters are derived from a short-horizon prediction of the \ac{VRU}'s motion, computed locally on the device using a lightweight state estimation algorithm. The resulting sequence of predicted states is then used to construct the ellipse representation that can be transmitted within the \ac{VAM}. The prediction method used to generate this motion intention is described in the next section.

\section{Path Prediction}
\label{sec:prediction}

To generate the intention representation described in the previous section, each \ac{VRU} computes a short-horizon prediction of its future motion based on recent trajectory observations. This prediction must be computationally lightweight to operate on resource-constrained on-board units embedded in bicycles or scooters while producing trajectory samples that can be compactly encoded for transmission in \acp{VAM}.

For this purpose, we employ an \ac{EKF} using a Constant Turn Rate and Velocity (CTRV) kinematic model. The \ac{EKF} is used to filter GNSS measurements and estimate the current motion state. Short-horizon trajectory prediction is then generated through open-loop propagation of the CTRV model using the most recent \ac{EKF} state estimate. 

The motion model is based on planar kinematics under the constant turn-rate and constant velocity assumption~\cite{Polychronopoulos2007}. Let $\vec{r}(t) \in \mathbb{R}^2$ and $\vec{v}(t) \in \mathbb{R}^2$ denote the position and velocity vectors, respectively. The dynamics are given by
\[
\dot{\vec{r}} = \vec{v}, \qquad
\dot{\vec{v}} = \omega J \vec{v}, \qquad
\dot{\omega} = 0,
\]
where
\[
J =
\begin{bmatrix}
0 & -1\\
1 & 0
\end{bmatrix}
\]
is the planar rotation generator and $\omega \in \mathbb{R}$ denotes the yaw rate.

Parameterizing the velocity in polar coordinates as $\vec{v} = \|\vec{v}\|[\cos\theta,\ \sin\theta]^{\mathsf{T}}$, and defining the speed as $v = \|\vec{v}\|$, the corresponding \ac{EKF} state vector is
\begin{equation}
\mathbf{x} =
\begin{bmatrix}
x & y & v & \theta & \omega
\end{bmatrix}^{\mathsf{T}}.
\end{equation}

Under the CTRV assumption, both the speed $v$ and yaw rate $\omega$ remain constant over the prediction interval, yielding the kinematic equations
\begin{equation}
\dot{x} = v \cos\theta, \quad
\dot{y} = v \sin\theta, \quad
\dot{\theta} = \omega, \quad
\dot{v} = 0, \quad
\dot{\omega} = 0.
\end{equation}

Integrating these equations over a discrete time interval $\Delta t$ yields the nonlinear CTRV transition model
\begin{equation}
\mathbf{x}_{k+1} = \mathbf{f}(\mathbf{x}_k)=
\begin{bmatrix}
x_k + \frac{v_k}{\omega_k}\left[\sin(\theta_k+\omega_k\Delta t)-\sin(\theta_k)\right] \\
y_k + \frac{v_k}{\omega_k}\left[-\cos(\theta_k+\omega_k\Delta t)+\cos(\theta_k)\right] \\
v_k \\
\theta_k + \omega_k \Delta t \\
\omega_k
\end{bmatrix}.
\end{equation}

The nonlinear transition function is linearized around the current state estimate to obtain the Jacobian
\[
F = \frac{\partial \mathbf{f}}{\partial \mathbf{x}},
\]
which is used to propagate the state covariance in the standard \ac{EKF} prediction step.

The predicted state sequence generated by this model is used in the next section to construct the uncertainty ellipse representation transmitted within the \acp{VAM}.

\section{Optimization of Predicted Trajectory Representations}
\label{sec:representation}

\subsection{Selection of the Optimal Representation Form}
\label{sec:selection}

\begin{figure}
    \centering
    \includegraphics[width=\linewidth, trim = {5cm 5.5cm 5.5cm 5.5cm}, clip]{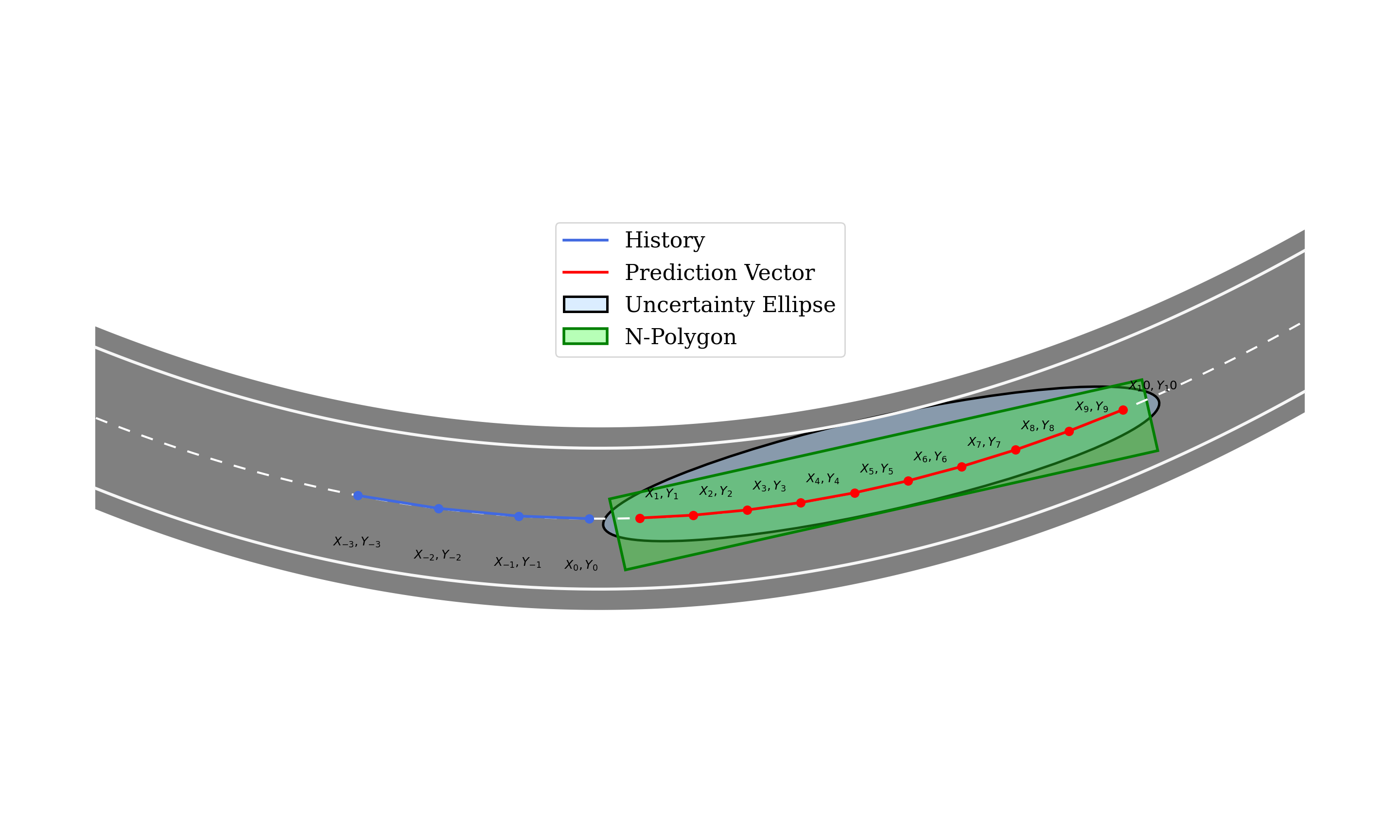}
    \caption{\ac{ETSI}-compliant Representation Forms}
    \label{fig:forms}
\end{figure}

After the open-loop \ac{EKF} prediction step generates a sequence of future position estimates, this information must be communicated to neighboring vehicles via \acp{VAM}. In accordance with \ac{ETSI} standards, predicted motion can be conveyed either explicitly as a trajectory vector or implicitly as a geographic area, such as an uncertainty ellipse or an $N$-polygon, as illustrated in Fig.~\ref{fig:forms}.

We argue that the uncertainty ellipse is particularly well suited for \ac{V2X} intention sharing for two main reasons. First, it provides a compact probabilistic description of the spatial dispersion of the predicted trajectory, allowing the maneuver to be summarized with a small set of parameters rather than a sequence of points. Second, our previous work~\cite{valle:hal-05424230} analytically showed that ellipse-based encoding achieves more favorable scalability bounds than trajectory vectors and $N$-polygon representations.

This probabilistic representation efficiently captures positional uncertainty, thereby enhancing robustness to measurement noise and prediction errors. Moreover, the communication overhead is significantly reduced compared with transmitting full trajectory vectors, as only the parameters of the covariance ellipse need to be shared. 

While the \ac{EKF} internally propagates a covariance matrix associated with the estimated motion state, this covariance primarily reflects local linearization uncertainty of the dynamic model. When predicting multiple steps ahead, nonlinear motion propagation and maneuver variability may cause the spatial dispersion of predicted states to deviate from this local approximation. For this reason, the uncertainty ellipse used for communication is derived from the empirical distribution of the predicted trajectory points. This approach captures the overall spatial spread of the predicted maneuver while remaining robust to modeling errors and nonlinear motion effects while adding minimal computational overhead.

Specifically, the predicted maneuver is encoded as a statistical confidence ellipse constructed from the sequence of predicted positions generated by the \ac{EKF} over the prediction horizon. Let ${p}_k =[{x}_k, {y}_k]^T$ denote the predicted position at step $k$, and let $p_{1},\ldots,p_{T}$ denote the predicted trajectory over the horizon $T$. The center of the ellipse is then defined as the sample mean of the predicted trajectory,

\begin{equation}
\boldsymbol{\mu} = \begin{bmatrix}
\mu_{x} \\ \mu_{y}
\end{bmatrix} =
\frac{1}{T}
\sum_{k=1}^{T} \begin{bmatrix} \mathbf{x}_k \\ \mathbf{y}_k \end{bmatrix}
\end{equation}

The spatial dispersion of the maneuver is captured by the sample covariance matrix,

\begin{equation}
\mathbf{\Sigma} =
\frac{1}{T-1}
\sum_{k=1}^{T}
(\mathbf{p}_k - \boldsymbol{\mu})
(\mathbf{p}_k - \boldsymbol{\mu})^\top.
\end{equation}

This formulation captures the spatial dispersion of the full predicted trajectory rather than only the local linearization uncertainty represented by the \ac{EKF} state covariance. The pair $(\mu,\Sigma)$ defines a bivariate normal distribution that approximates the predicted motion region.

\begin{equation}
\label{eq:region}
(\mathbf{p} - \boldsymbol{\mu})^\top
\mathbf{\Sigma}^{-1}
(\mathbf{p} - \boldsymbol{\mu})
\le \chi^2_{2,\alpha},
\end{equation}

where $p \in \mathbb{R}^2$ denotes an arbitrary spatial position and $\chi^2_{2,\alpha}$ is the chi-square quantile with two degrees of freedom corresponding to confidence level $\alpha$. 

The maneuver can therefore be fully described by five floating-point parameters: the mean position $(\mu_x,\mu_y)$ and the three independent elements of the symmetric covariance matrix $\mathbf{\Sigma}$. 
This formulation preserves statistical interpretability while maintaining a constant payload size independent of the number of predicted trajectory samples.

\subsection{Collision Detection Using Ellipse Representations}
\label{collisondetec}

Once motion intentions are exchanged through \acp{VAM},\acp{VRU} can use the transmitted maneuver representations to assess potential conflicts with their own predicted trajectories. This collision assessment constitutes the next step in the intention sharing pipeline, enabling cooperative maneuver coordination and collision avoidance. While the present work focuses on the scalability of trajectory representation and intention sharing, the uncertainty ellipse encoding naturally supports efficient geometric collision checks. For completeness, we briefly outline how collision detection between two predicted motion regions can be formulated as a convex feasibility problem. A detailed evaluation of cooperative collision avoidance strategies is beyond the scope of this paper.

To directly test whether two ellipsoids intersect, the problem can be formulated as a nonconvex \ac{QCQP}. As previously shown, given two multivariate Gaussian distributions $\mathcal{N}(\bm{\mu}_1, \bm{\Sigma}_1)$ and $\mathcal{N}(\bm{\mu}_2, \bm{\Sigma}_2)$ in $\mathbb{R}^d$, each distribution defines an ellipsoidal confidence region:

\begin{equation}
    \mathcal{E}_i = \left\{ \mathbf{x} \in \mathbb{R}^d \mid (\mathbf{x} - \bm{\mu}_i)^\top \bm{\Sigma}_i^{-1} (\mathbf{x} - \bm{\mu}_i) \leq c_i \right\},
\end{equation}

where \( \bm{\mu}_i \in \mathbb{R}^d \) is the mean, \(\bm{\Sigma}_i \in \mathbb{R}^{d \times d}\) is the covariance matrix, and \( c_i = \chi^2_{d, \alpha} \) is the chi-squared quantile at confidence level \(\alpha\), which can be equivalently represented by a quadratic form

\begin{align}
    \mathcal{E}_i &= \left\{ \mathbf{x} \in \mathbb{R}^d \mid f_i(\mathbf{x}) = \mathbf{x}^\top \mathbf{A}_i \mathbf{x} + 2\mathbf{b}_i^\top \mathbf{x} + c_i^{(\mathrm{quad})} \leq 0 \right\} 
\end{align}

where $\mathbf{A}_i = \bm{\Sigma}_i^{-1}$, $ \mathbf{b}_i = -\mathbf{A}_i \bm{\mu}_i$ and $c_i^{(\mathrm{quad})} = \bm{\mu}_i^\top \mathbf{A}_i \bm{\mu}_i - c_i $.

The goal is to determine whether the intersection of two ellipses \( \mathcal{E}_1 \cap \mathcal{E}_2 \) is nonempty. This leads to the feasibility \ac{QCQP}:
\begin{equation}\label{eq:qcqp}
    \begin{aligned}
        \text{Find} \quad &\mathbf{x} \in \mathbb{R}^d \\
        \text{s.t.} \quad &f_1(\mathbf{x}) \leq 0, \\
                          &f_2(\mathbf{x}) \leq 0.
    \end{aligned}
\end{equation}

If a feasible \( \mathbf{x} \) exists, the ellipsoids intersect. Otherwise, they are disjoint.



\subsection{Complexity of Collision Detection in 2D}
\label{complexity}

In \ac{V2X} scenarios, uncertainty regions associated with \ac{VRU} trajectories are modeled as confidence ellipses derived from positive definite covariance matrices, ensuring strictly convex quadratic forms \(\mathbf{A}_i \succ 0\). Under this assumption, each ellipsoid defines a convex set, and the intersection test can be formulated as a small convex quadratic feasibility problem. In \(\mathbb{R}^2\), this formulation involves only two decision variables, resulting in fixed and very low dimensionality.

Such convex \acp{QCQP} can be efficiently solved using interior-point methods, which have worst-case complexity \(\mathcal{O}(n^3 \log(1/\epsilon))\), where \(n=2\) is the number of variables and \(\epsilon\) the desired numerical precision~\cite{nesterov1994}. In this setting, each Newton iteration requires solving a \(2\times2\) linear system,  which is computationally trivial.

Compared with explicit trajectory-vector representations, where collision checks require evaluating intersections between sequences of sampled trajectory points and therefore scale linearly with the prediction horizon $O(T)$, the ellipse-based formulation enables collision assessment in constant time $O(1)$, independent of the number of predicted trajectory samples. This property makes ellipse-based representations particularly attractive for real-time intention sharing in dense V2X environments, where multiple neighboring trajectories must be evaluated simultaneously.

\section{Evaluation}
\label{sec:evaluation}

\subsection{Simulation Scenario}

\acp{CAM} and \ac{VAM} exchanges are simulated using the Artery framework \cite{artery}, which extends the Veins vehicular networking simulator with \ac{ETSI}-compliant support for the \ac{VBS}. The scenario models vehicles and cyclists operating in a Manhattan-style urban grid environment under dense urban traffic conditions.

\acp{VAM} are generated according to the \ac{ETSI} specification and periodically transmit motion prediction containers containing the predicted trajectories produced by the \ac{EKF} pipeline. Vehicles simultaneously broadcast \acp{CAM} while receiving \acp{VAM}, thereby maintaining realistic background channel load and cooperative awareness dynamics throughout the simulation.

To evaluate the scalability of the proposed intention sharing framework under dense \ac{VRU} conditions, the traffic scenario consists of four flows traveling west-to-east along a central arterial and splitting midway toward the northeast and southeast. Each flow contains either 200 vehicles or 300 bicycles. Bicycles are injected at a rate ten times higher than vehicles in order to emulate dense \ac{VRU} traffic and stress communication and computational performance. Each configuration is repeated five times for every trajectory representation form, and the reported results correspond to the mean values across all runs.

\subsection{Simulation-Based Scalability Analysis}

\begin{table*}[htbp]
\centering
\caption{Estimated Normalized CPU Clock Cycles for Intention Sharing under Different Representation Forms ($T=15$, $V=4$, $C=1$). 
All values are expressed as factors of $10^3$.}
\label{tab:cpu_operations_is}

\begin{tabular}{ccccc}
\toprule
Distance Bin & Average Neighbors & Uncertainty Ellipse & N-Polygon & Trajectory Vector \\
\midrule
$[0,100]$   & 134.96 & 18.22 & 72.89  & 273.34 \\
$(100,200]$ & 164.33 & 27.01 & 108.04 & 405.14 \\
$(200,300]$ & 187.97 & 35.33 & 141.32 & 529.56 \\
$(300,400]$ & 206.78 & 42.76 & 171.04 & 641.38 \\
$(400,500]$ & 222.90 & 49.68 & 198.71 & 745.86 \\
\bottomrule
\end{tabular}

\end{table*}

To analyze how computational effort scales with the number of neighboring road users, simulation results are grouped according to distance bins representing increasing communication ranges around each ego agent. For each bin, the corresponding average number of neighbors is computed and used to evaluate the computational cost of the different trajectory representations.


Table~\ref{tab:cpu_operations_is} reports the estimated normalized CPU clock cycle counts for \textit{intention sharing}, including both prediction and collision checking, using three \ac{ETSI}-compliant representations. The trajectory-vector results assume a linear-in-$T$ model ($T=15$, the maximum number of points allowed for transmission), where matching is performed using precomputed offsets rather than full point-wise comparisons. For the polygon representation, a configuration with $V=4$ vertices is adopted in accordance with \cite{etsi_geo_area_2025}, providing a suitable trade-off between geometric fidelity and computational cost while introducing only a linear factor in complexity.

CPU clock cycle counts are used as a hardware independent proxy for computational effort, enabling fair comparison between representations while avoiding dependencies on compiler optimizations or runtime environments. Cycle counts are obtained during simulation using the \texttt{rdtsc} instruction to measure processor cycle counts in the prediction–encoding and decoding–collision checking blocks. This approach provides a consistent low-level estimate of computational cost across representation methods.

\begin{figure}[t]
    \centering
    \includegraphics[width=\linewidth]{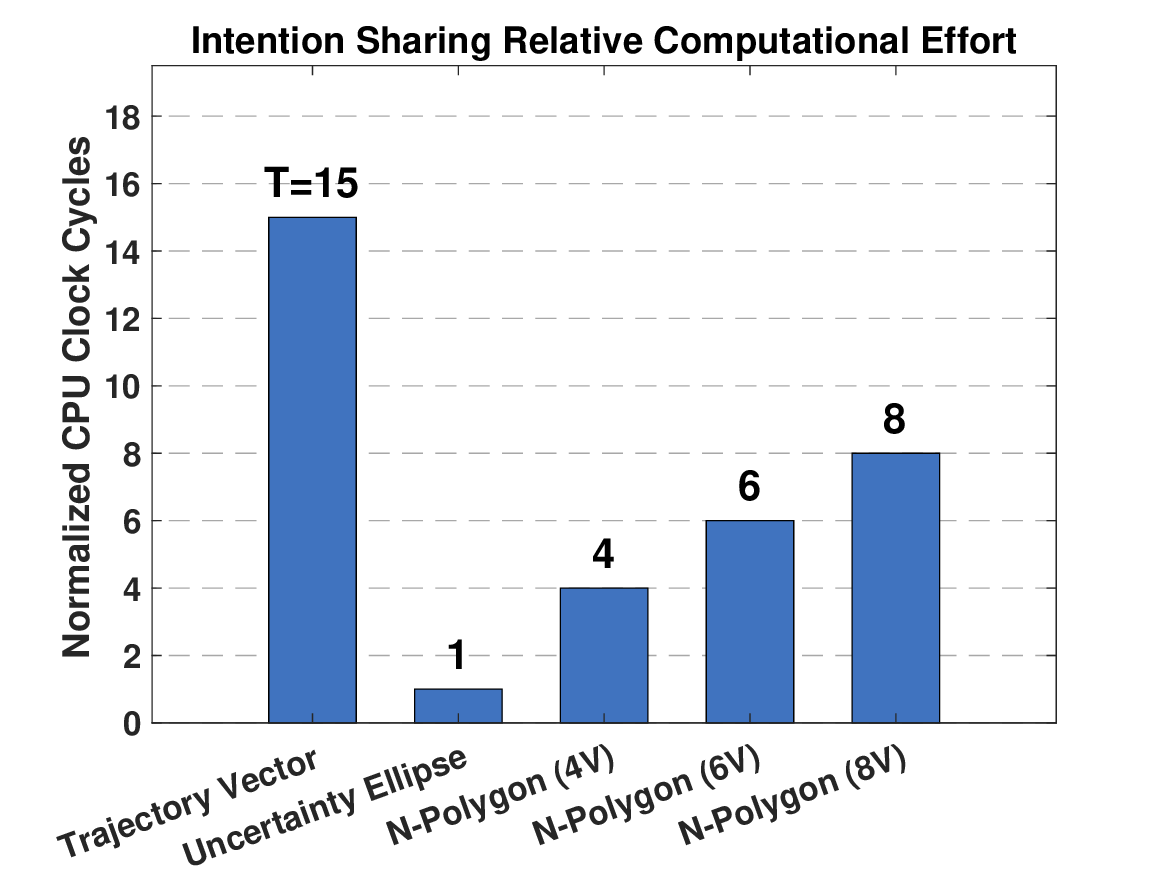}
    \caption{Relative analytical computational effort required for intention sharing using different trajectory representations}
    \label{fig:effort}
\end{figure}

The relative computational effort across the representation forms follows the analytical complexity trends summarized in Fig.~\ref{fig:effort}. In particular, the measured results confirm that the trajectory-vector representation scales with both the neighborhood size and prediction horizon, whereas both the uncertainty-ellipse and the $N$-polygon representations remove the dependence on the prediction horizon, validating the theoretical scalability advantages derived in \cite{valle:hal-05424230}. 

Simulation results indicate that for neighborhood sizes in the range $N \approx 135$--$223$, trajectory-vector representations require on the order of $10^{5}$--$10^{6}$ CPU clock cycles per simulation under the linear-$T$ model. In comparison, $N$-polygons ($V=4$) require roughly $10^{5}$ operations, whereas uncertainty ellipses incur the lowest cost at approximately $10^{4}$ operations. These results confirm that the quadratic dependence on neighborhood size dominates overall runtime, while the trajectory representation determines the multiplicative scaling factor.

From a practical standpoint, implicit geometric representations such as uncertainty ellipses or low-vertex $N$-polygons provide superior scalability for real-time intention sharing in dense traffic scenarios. By eliminating the dependence on the prediction horizon $T$, they maintain significantly lower computational loads than explicit trajectory vectors. Although trajectory-vector representations capture richer temporal detail, their computational burden grows rapidly, especially if full point-wise matching is required. Offset-based matching can mitigate this cost in low-density networks, but for large-scale cooperative perception and \ac{V2X} systems, combining intention sharing with implicit uncertainty representations remains the most computationally efficient configuration.

\subsection{Field Data Collection}

\begin{figure}
     \centering
     \includegraphics[width=1.0\linewidth]{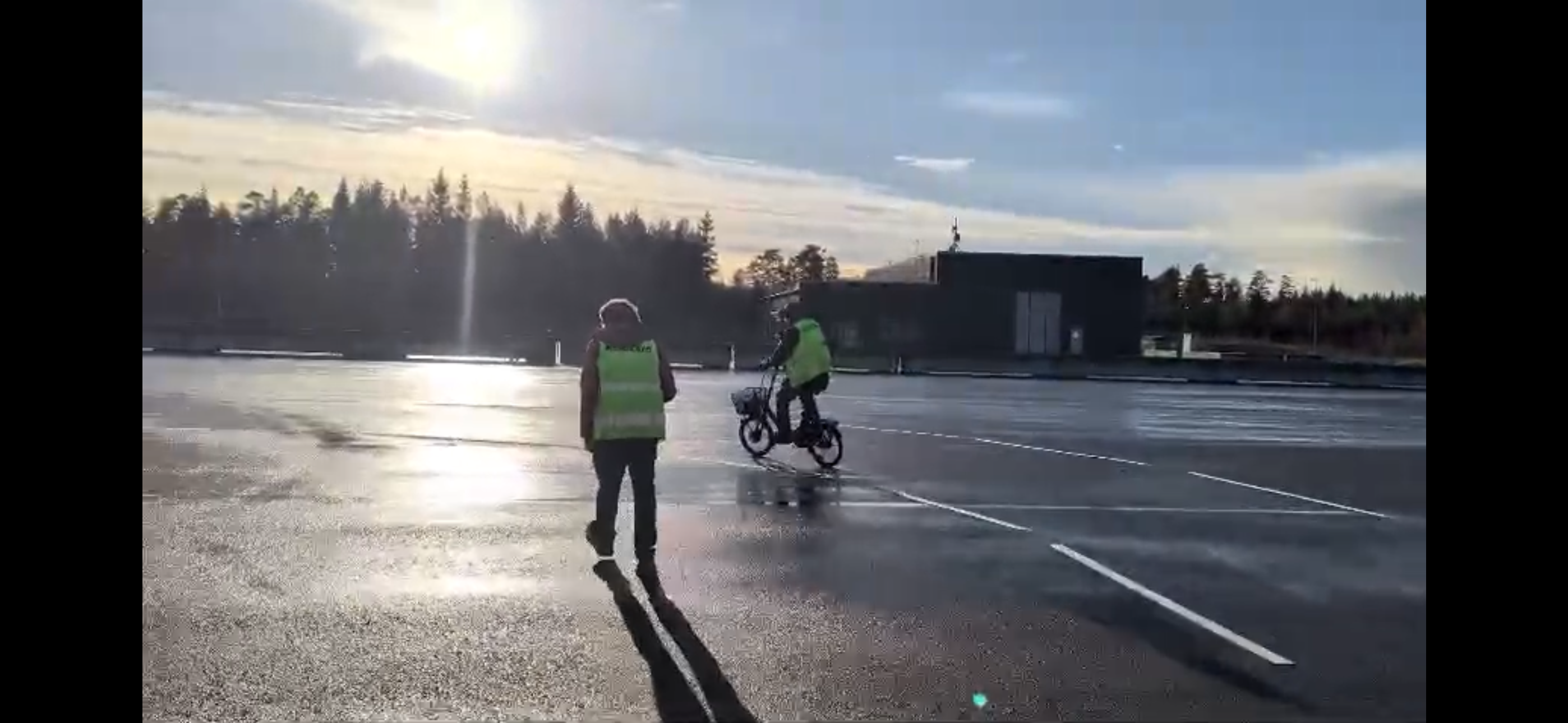}
     \caption{Left turn maneuver data collection at AstaZero}
     \label{fig:pic-az}
 \end{figure}

Field experiments were conducted at the AstaZero Test Facility in Borås, Sweden to evaluate the EKF prediction performance. Using an instrumented electric bicycle, GNSS measurements were collected for a range of maneuvers, including typical scenarios (e.g., left-hand turns) and highly non-stationary behaviors (e.g., randomized zig-zag trajectories) to probe the operational limits of the prediction algorithm. Fig.~\ref{fig:pic-az} shows the collection for a left turn maneuver under controlled conditions (i.e., a single bicycle and no vehicular traffic).

Position and timing data were obtained using V2X equipment with 3D GNSS capabilities integrated into an on-board unit mounted on the bicycle. In addition, a Raspberry Pi running our protocol collected GNSS data from an independent USB-connected GNSS receiver for comparison and redundancy.


\begin{figure}[t]
  \centering
  \includegraphics[width=\linewidth]{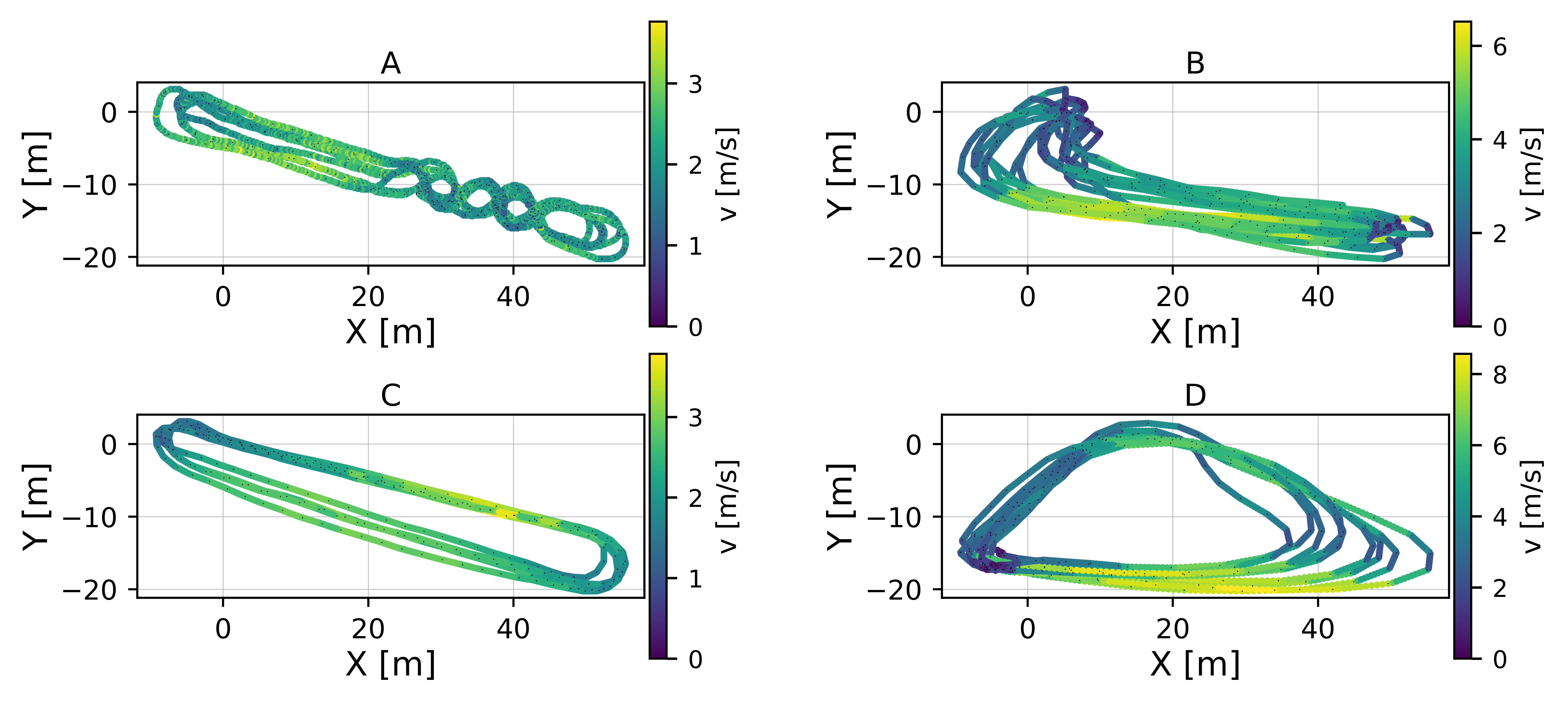}
  \caption{Sample Test Track Maneuvers}
  \label{fig:gps_maneuvers}
\end{figure}

\subsection{EKF Prediction Performance}

This section evaluates the \ac{EKF} for short-horizon open-loop prediction under both nominal and highly non-stationary bicycle maneuvers. We focus on usable prediction horizons and covariance propagation, which directly determines the uncertainty ellipse transmitted within \ac{ETSI} \acp{VAM}. For reference, the EKF predictor is compared with an \ac{EKF} with Constant Velocity (CV) model and the polynomial \ac{LSM} method used in our previous work.

\subsubsection{EKF Prediction Block - Uncertainty Representation}

We first evaluate the proposed \ac{EKF} predictor under highly non-stationary motion patterns in order to assess its robustness and operational limits. In particular, trajectories with abrupt variations, such as the one shown in Fig.~\ref{fig:gps_maneuvers} A, are used as stress cases for rapidly changing cyclist motion. Although such maneuvers are unlikely under typical road conditions, they are useful for analyzing how quickly the predictor adapts to sudden changes in motion intent and how this impacts the motion information shared through \acp{VAM}.

\begin{figure}[t]
  \centering
  \includegraphics[width=\linewidth, trim = {5cm 0.5cm 5.5cm 2cm}, clip]{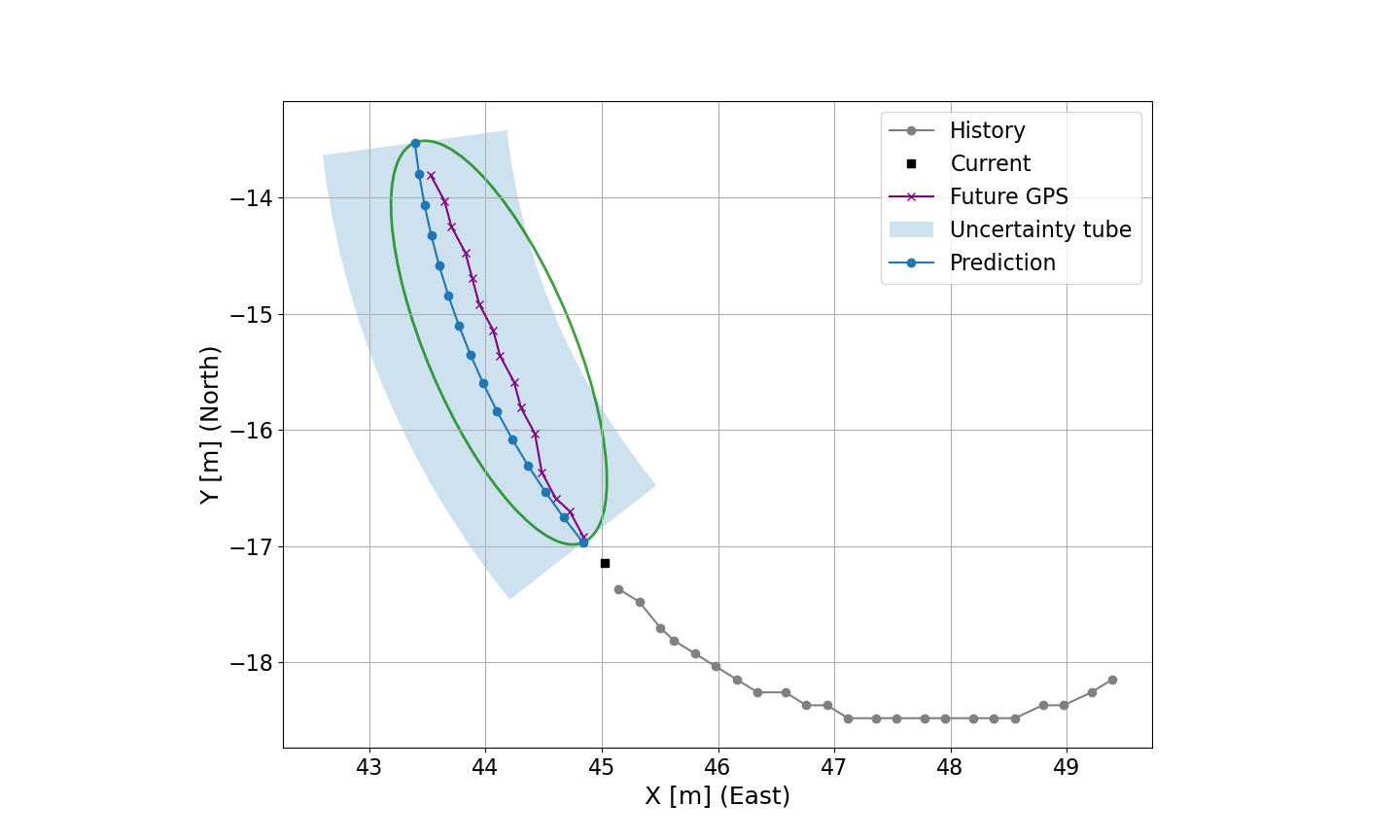}
  \caption{Example EKF Prediction Block}
  \label{fig:ekf_block}
\end{figure}

Fig.~\ref{fig:ekf_block} illustrates an example prediction block generated by the \ac{EKF} with a horizon of $T = 1.5\,\mathrm{s}$ ($15$ discrete steps). According to the \ac{ETSI} specification, the path prediction container can transmit either up to $15$ predicted points or $5\,\mathrm{s}$ of future intention, depending on the sampling interval, whichever limit is reached first. The predicted states closely follow the measured GNSS positions, while the associated uncertainty ellipse represents the dispersion of the predicted trajectory samples for transmission within the corresponding \ac{VAM}. Small discrepancies between predictions and measurements mainly reflect the locally linearized motion model of the \ac{EKF}, which produces a smooth estimate, whereas the measurements include sensor noise and minor motion irregularities.

Because the ellipse captures the spatial dispersion of the predicted states, it can also be mapped to a prediction tube (uncertainty corridor), as can also be seen in Fig.~\ref{fig:ekf_block}. This visualization provides an intuitive view of covariance propagation along the predicted maneuver relative to typical bicycle-lane boundaries. Although primarily illustrative, it highlights how ellipse parameters can define a corridor-level safety region while enabling compact transmission of both predicted motion and uncertainty, avoiding the overhead of full trajectory vector encoding.

\subsubsection{Prediction Performance Under Relaxed Error Thresholds}

\begin{table}[t]
\centering
\caption{Prediction Accuracy and Usable Horizon Statistics}
\label{tab:usable_horizon_stats}
\setlength{\tabcolsep}{3pt}
\footnotesize
\begin{tabular}{ccccc}
\toprule
& \multicolumn{2}{c}{Non-Stationary} & \multicolumn{2}{c}{Left Turn} \\
\cmidrule(lr){2-3} \cmidrule(lr){4-5}
$\epsilon$ & RMSE (m) & Horizon (s) & RMSE (m) & Horizon (s) \\
\midrule
1.0 & $0.903\!\pm\!0.008$ & $0.90\!\pm\!0.00$ & $0.962\!\pm\!0.035$ & $1.80\!\pm\!0.076$ \\
2.0 & $1.899\!\pm\!0.006$ & $1.70\!\pm\!0.00$ & $1.934\!\pm\!0.050$ & $3.30\!\pm\!0.106$ \\
4.0 & $3.912\!\pm\!0.083$ & $2.90\!\pm\!0.053$ & $3.947\!\pm\!0.217$ & $5.80\!\pm\!0.001$ \\
6.0 & $5.950\!\pm\!0.078$ & $3.90\!\pm\!0.194$ & $4.036\!\pm\!0.217$ & $5.90\!\pm\!0.002$ \\
\bottomrule
\end{tabular}
\end{table}


Table~\ref{tab:usable_horizon_stats} summarizes the prediction accuracy of the proposed \ac{EKF}-based open-loop predictor evaluated at its usable horizon under progressively relaxed spatial error thresholds~$\epsilon$, for both a highly non-stationary stress-test trajectory and a realistic left-hand turning maneuver. Rather than reporting block-averaged RMSE over a fixed horizon, the usable horizon metric captures the maximum prediction duration for which the average positional error remains below a specified tolerance. This formulation aligns well with intention sharing, where conveying the overall motion tendency is more important than maintaining point-wise positional accuracy at long horizons.

For the highly non-stationary trajectory shown in Fig.~\ref{fig:gps_maneuvers} A, strict accuracy requirements still limit predictive utility due to frequent curvature reversals and speed fluctuations. At $\epsilon = 1$~m the usable horizon is approximately $0.9$~s, reflecting the difficulty of extrapolating rapidly changing motion. As the tolerance increases, the usable horizon grows steadily, reaching approximately $2.9$~s at $\epsilon = 4$~m and extending to nearly $4$~s at $\epsilon = 6$~m.

In contrast, the nominal left-hand turn maneuver (Fig.~\ref{fig:gps_maneuvers} B) exhibits substantially more stable behavior. At $\epsilon = 1$~m the usable horizon already reaches approximately $1.8$~s while maintaining sub-meter RMSE. Increasing the tolerance to $\epsilon = 2$~m extends the horizon to more than $3$~s, and for $\epsilon \geq 4$~m the usable horizon approaches $6$~s while maintaining prediction errors below approximately $4$~m.

The standard deviation values in Table~\ref{tab:usable_horizon_stats} remain small for both trajectory types, indicating consistent usable horizons and RMSE across repeated runs. While the mean prediction horizons differ substantially between the non-stationary and left-hand turn maneuvers, the variability itself is limited in both cases, suggesting stable EKF behavior under both motion patterns.

Overall, the results indicate that the reduced usable horizons in the stress-test scenario stem primarily from the intrinsic unpredictability of highly irregular motion rather than limitations of the estimator. For realistic cyclist maneuvers, the \ac{EKF} predictor provides stable multi-second prediction horizons that are sufficient for intention sharing within \ac{VAM} constraints.

\subsubsection{Comparison with Baseline Prediction Models}

\begin{table*}[t]
\centering
\caption{Mean usable prediction horizon (s) and RMSE at the horizon under different error thresholds.}
\label{tab:predictor_comparison_summary}
\begin{tabular}{ccccccc}
\toprule
$\boldsymbol{\epsilon}$ (m) 
& \multicolumn{2}{c}{\textbf{EKF-CTRV}} 
& \multicolumn{2}{c}{\textbf{EKF-CV}} 
& \multicolumn{2}{c}{\textbf{Poly-LSM (deg=2)}} \\
\cmidrule(lr){2-3} \cmidrule(lr){4-5} \cmidrule(lr){6-7}
& Horizon (s) & RMSE (m) 
& Horizon (s) & RMSE (m) 
& Horizon (s) & RMSE (m) \\
\midrule
1.0 & $1.20 \pm 0.10$ & $0.96 \pm 0.02$ & $0.39 \pm 0.33$ & $0.96 \pm 0.03$ & $0.37 \pm 0.46$ & $0.94 \pm 0.05$ \\
2.0 & $2.14 \pm 0.16$ & $1.95 \pm 0.03$ & $1.19 \pm 0.48$ & $1.94 \pm 0.04$ & $0.79 \pm 0.73$ & $1.92 \pm 0.05$ \\
4.0 & $3.58 \pm 0.29$ & $3.92 \pm 0.05$ & $2.55 \pm 0.54$ & $3.92 \pm 0.05$ & $1.50 \pm 0.97$ & $3.84 \pm 0.09$ \\
6.0 & $4.71 \pm 0.39$ & $5.88 \pm 0.12$ & $3.67 \pm 0.65$ & $5.90 \pm 0.06$ & $2.05 \pm 1.09$ & $5.86 \pm 0.14$ \\
\bottomrule
\end{tabular}
\end{table*}

\begin{figure}[t]
  \centering
  \includegraphics[width=\linewidth]{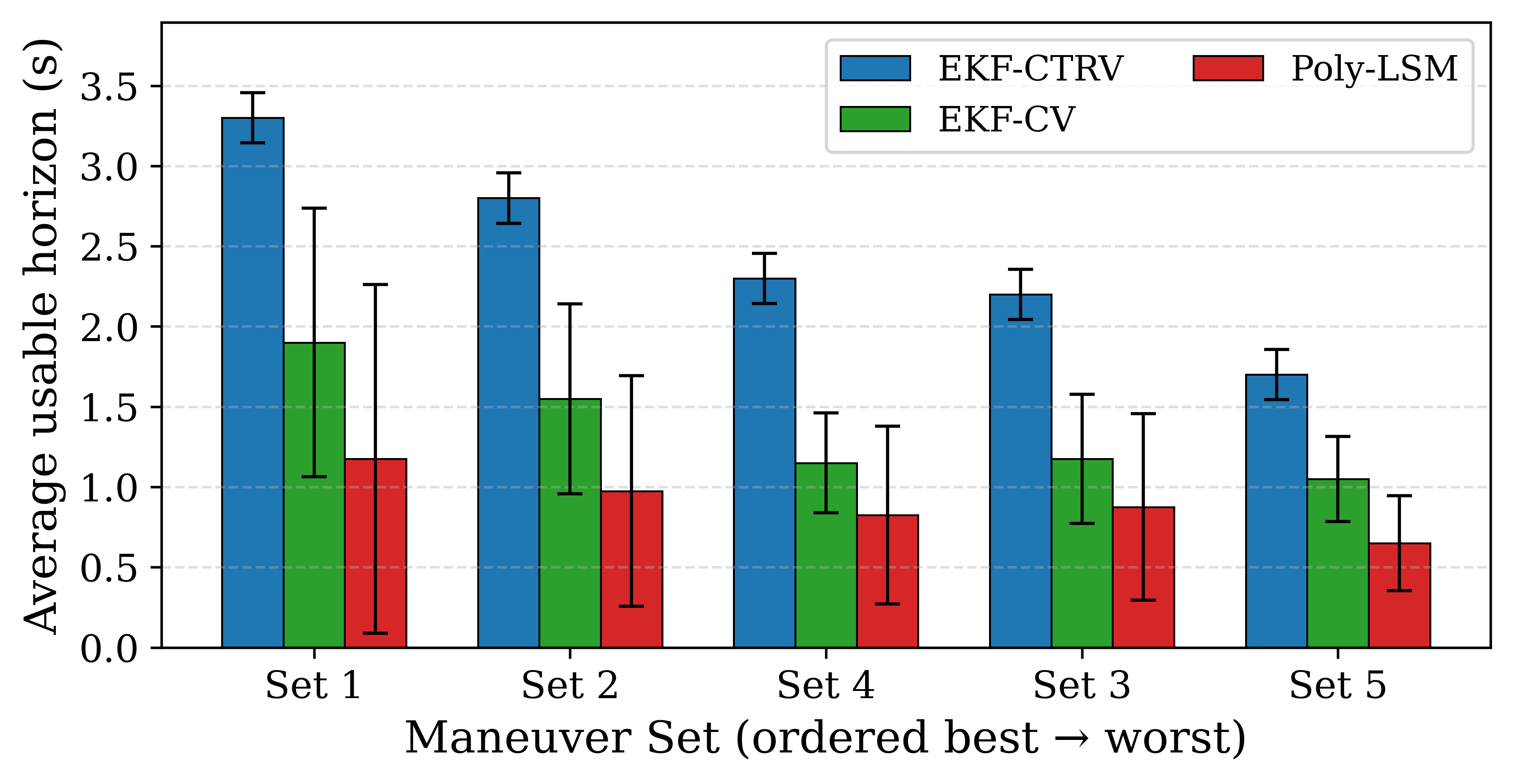}
  \caption{Average Usable Horizons per Maneuver Set for $\boldsymbol{\epsilon} = 2m$}
  \label{fig:horizon_per_maneuver}
\end{figure}

To contextualize the performance of the proposed \ac{EKF}-CTRV predictor, we compare it with two baseline model: an \ac{EKF} using a Constant Velocity (CV) model and the polynomial \ac{LSM} predictor used in our previous work \cite{valle:hal-05424230}. The \ac{EKF}-CV model assumes straight-line motion with constant speed, whereas the \ac{LSM} predictor extrapolates a second-order polynomial fitted to recent trajectory samples.

Table~\ref{tab:predictor_comparison_summary} summarizes the mean usable prediction horizons and corresponding RMSE values across the complete set of test-track trajectories. While all predictors achieve comparable RMSE values at the evaluation thresholds, clear differences appear in the usable horizons. The \ac{EKF}-CTRV model consistently yields the longest horizons, reflecting its ability to capture curvature dynamics typical of bicycle maneuvers. The \ac{EKF}-CV baseline performs similarly in near-linear motion but degrades when sustained heading changes occur. In contrast, the \ac{LSM} predictor produces the shortest horizons due to the rapid divergence of polynomial extrapolation under changing curvature.

To examine predictor behavior across motion types, Fig.~\ref{fig:horizon_per_maneuver} compares usable horizons for each maneuver set, including left-hand turns, start–stop motion, smooth circular trajectories, and zig-zag patterns. The maneuvers are ordered by increasing prediction difficulty, reflected by decreasing average usable horizons. Across all maneuver types, the \ac{EKF}-CTRV predictor maintains the most stable horizons, while the \ac{EKF}-CV model degrades for turning behaviors and the \ac{LSM} baseline shows the largest variability.

Overall, incorporating curvature dynamics through the CTRV model significantly improves prediction robustness for cyclist trajectories. The \ac{EKF}-CTRV predictor achieves longer usable horizons across diverse maneuvers while maintaining comparable positional accuracy, making it well suited for intention sharing where conveying motion tendency over multi-second horizons is more valuable than minimizing instantaneous positional error.

\section{Conclusion}
\label{sec:conclusion}

This work investigated scalable representations for maneuver intention sharing in \ac{V2X} systems. Three \ac{ETSI}-compliant geometric encodings—trajectory vectors, uncertainty ellipses, and $N$-polygons—were analyzed with respect to computational complexity and communication overhead.

Simulation-based CPU cycle measurements confirmed the analytical trends, showing that uncertainty ellipses maintain constant payload size while exhibiting favorable quadratic scaling with neighborhood size. Compared with vectors and $N$-polygons, the ellipse representation significantly reduces computational overhead in dense traffic scenarios.

To evaluate predictive feasibility, an \ac{EKF}-based open-loop predictor was assessed using a usable-horizon metric defined by a spatial tolerance $\epsilon$. Results indicate that reduced horizons in stress-test scenarios arise primarily from intrinsic motion unpredictability rather than estimator limitations. For realistic bicycle maneuvers, usable prediction horizons exceeding two seconds were achieved, which is sufficient for practical intention sharing within \ac{ETSI} \ac{VAM} constraints.

Overall, combining \ac{EKF}-based motion prediction with uncertainty ellipse encoding enables efficient and scalable communication of maneuver intent for \acp{VRU}. This approach provides a practical balance between prediction accuracy, communication efficiency, and computational scalability for cooperative V2X systems. Future work will investigate adaptive prediction horizons and context-aware message triggering strategies to further improve robustness and communication efficiency in highly dynamic traffic environments.

\bibliographystyle{IEEEtran}
\bibliography{references}

\begin{IEEEbiography}
[{\includegraphics[width=1in,height=1.25in,clip,keepaspectratio]{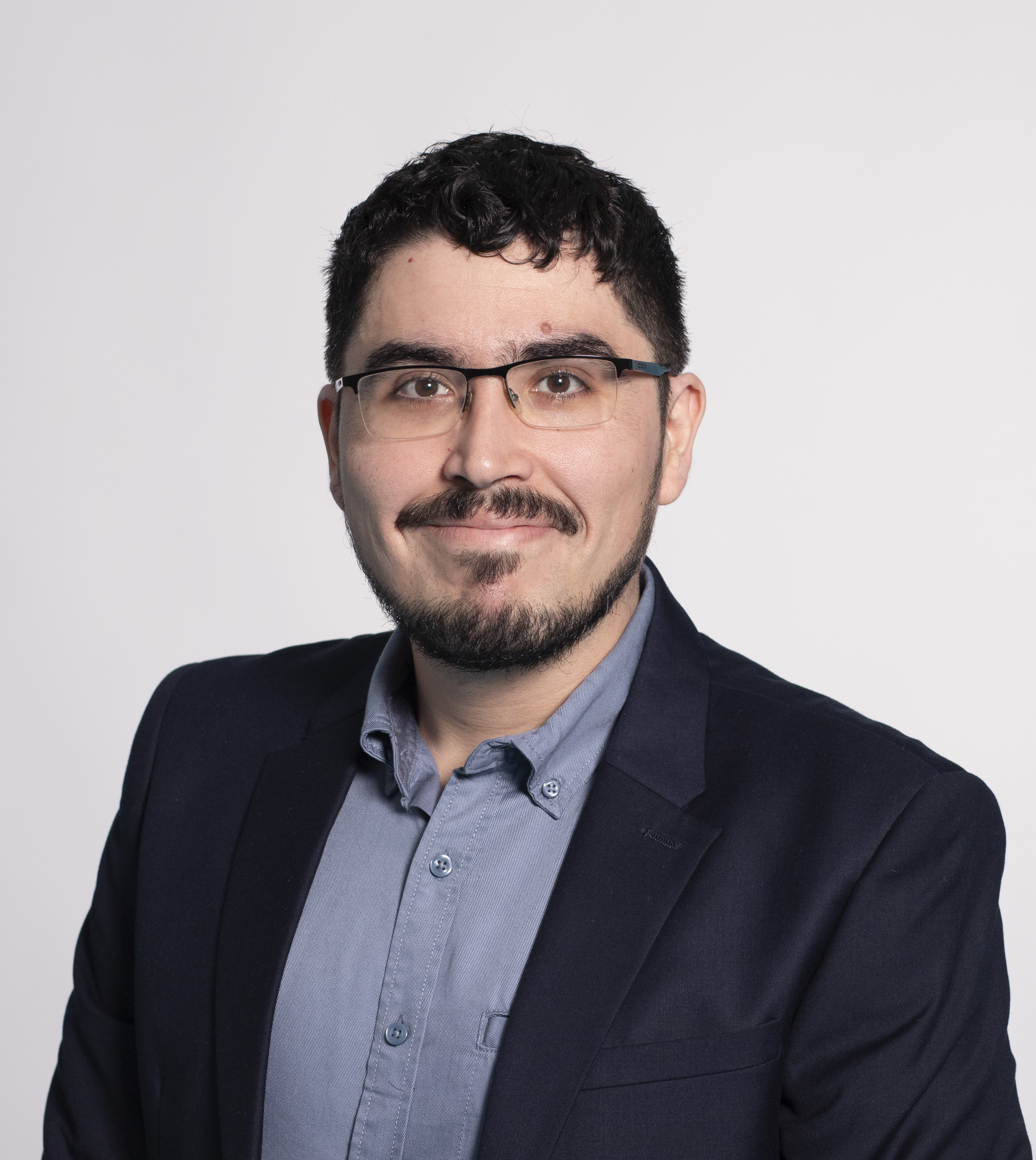}}]{Felipe Valle } received the M.Sc. degree in electrical engineering from the University of Chile, in 2019. He is currently pursuing a Ph.D. degree with Halmstad University, Sweden. His research interests include cooperative automated and autonomous driving systems, vehicular networking and vulnerable road user safety.
\end{IEEEbiography}

\begin{IEEEbiography}
[{\includegraphics[width=1in,height=1.25in,clip,keepaspectratio]{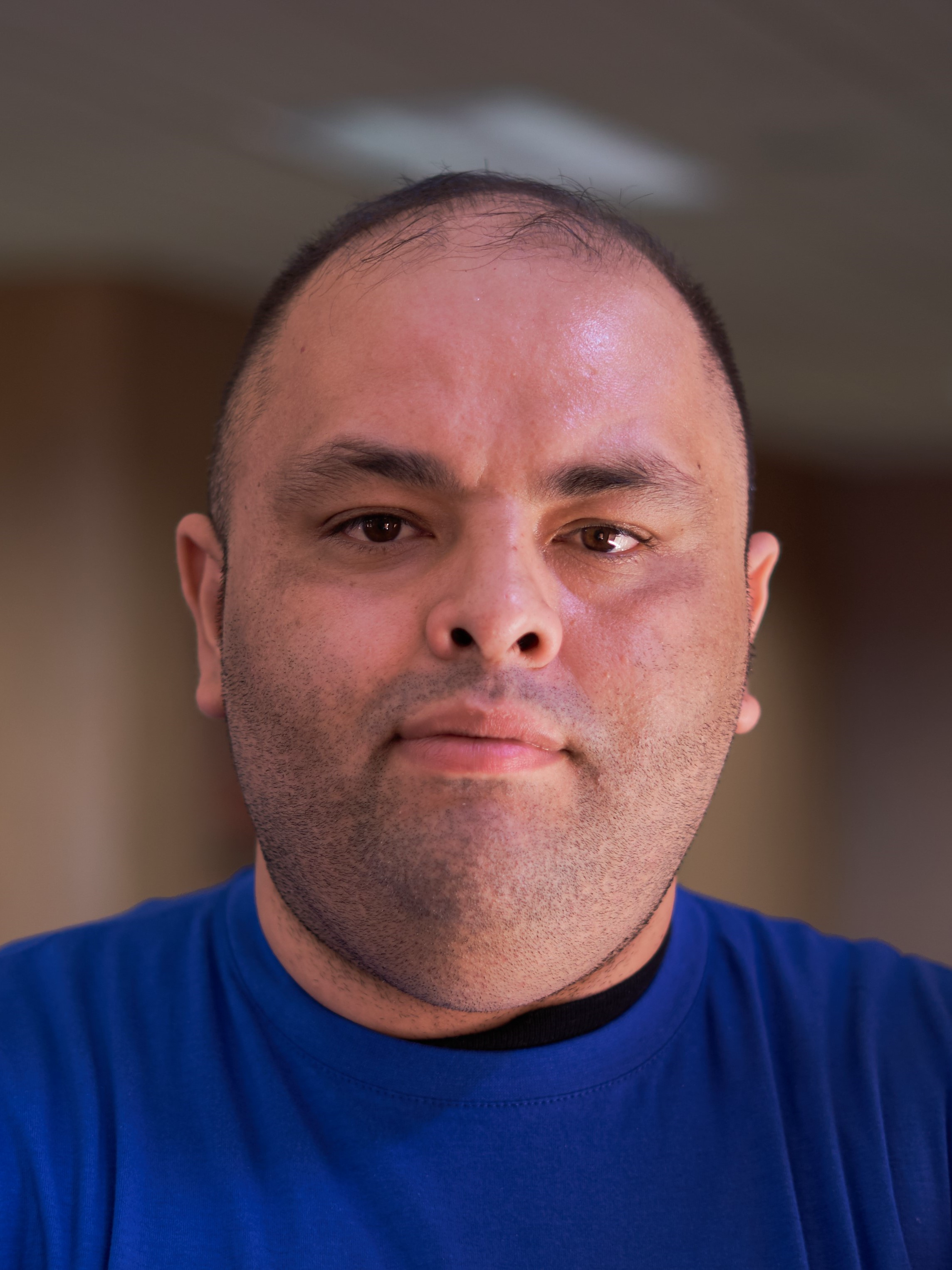}}]{Oscar Amador } received the bachelor’s degree in Telematics engineering from the Universidad Politécnica de Durango, Mexico (2012), and the M.Sc. degree in Telematics engineering from the University Carlos III of Madrid, Spain, sponsored by the Fundación Carolina Scholarship (2016), and the Ph.D. degree at University Carlos III of Madrid (2020). He is currently an Assistant Professor at Halmstad University, Sweden. His research interests are focused on vehicular networking, protection of vulnerable road users, and intelligent transport systems. 
\end{IEEEbiography}

\begin{IEEEbiography}
[{\includegraphics[width=1in,height=1.25in,clip,keepaspectratio]{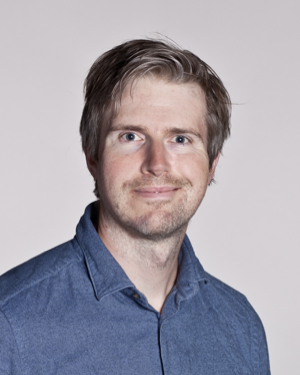}}]{Johan Thunberg } is a WASP Fellow and an Associate Professor at the Department of Electrical and Information Technology at Lund University, Sweden. He received his Ph.D. degree from the KTH Royal Institute of Technology, Sweden, in 2014. Between 2014 and 2018, he held a position as a Research Associate with the Luxembourg Centre for Systems Biomedicine, University of Luxembourg, Luxembourg. Between 2018 and 2023 was with the School of Information Technology, Halmstad University, first as Assistant Professor and subsequently as Associate Professor. His research interests covers communications, control theory, nonlinear systems, and pattern recognition. 
\end{IEEEbiography}

\begin{IEEEbiography}
[{\includegraphics[width=1in,height=1.25in,clip,keepaspectratio]{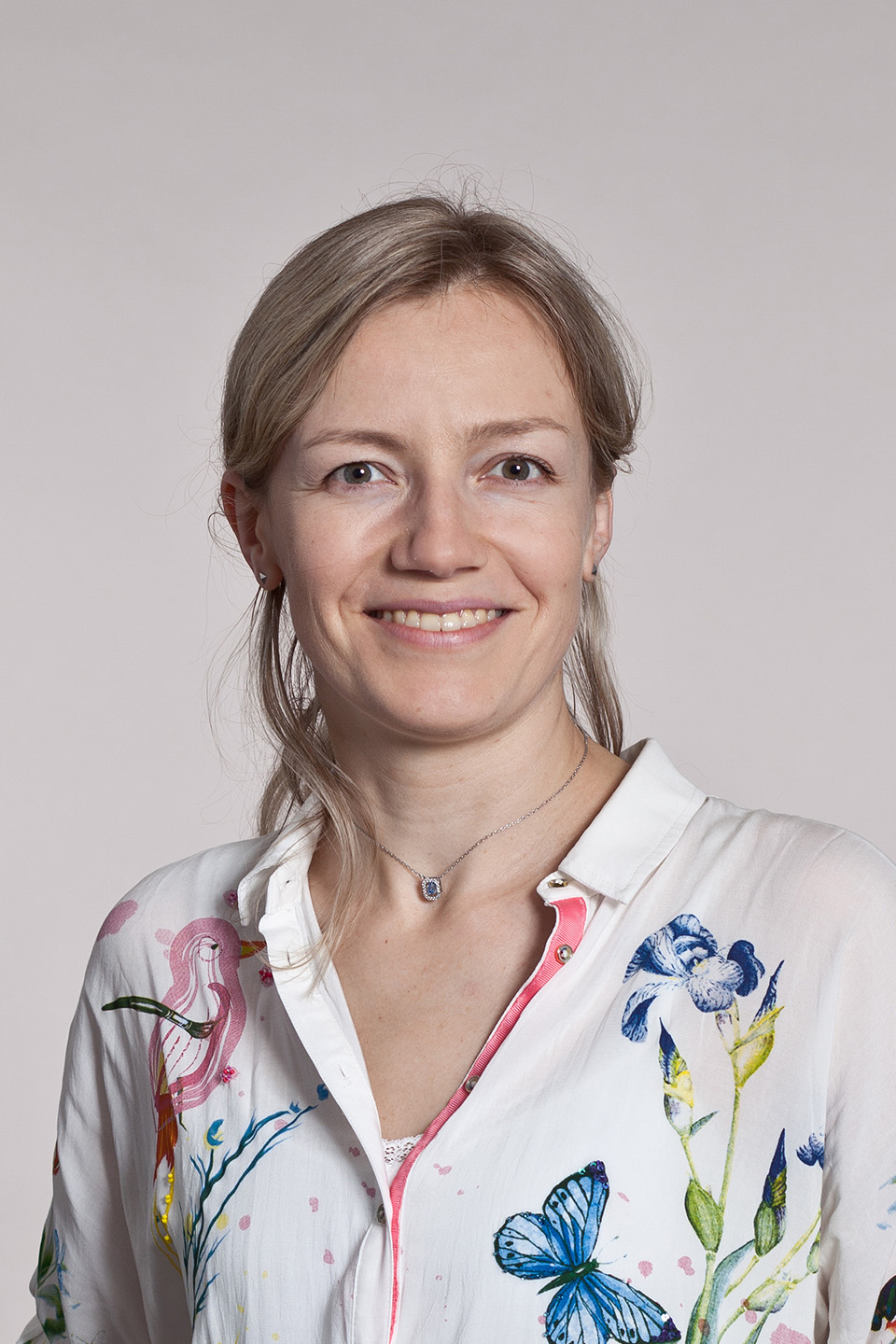}}]{Elena Haller } received the Ph.D. degree in applied mathematics from Luleå University of Technology, Sweden, in 2021. She is currently a Senior Lecturer in Applied Mathematics with Halmstad University, Sweden. Her research interests include fluid dynamics, traffic flow modeling, and mathematical modeling for mobility and transportation systems.
\end{IEEEbiography}

\begin{IEEEbiography}
[{\includegraphics[width=1in,height=1.25in,clip,keepaspectratio]{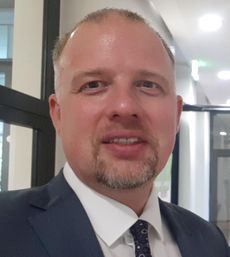}}]{Alexey Vinel }  is a professor at the Karlsruhe Institute of Technology (KIT), Germany. Previously he was a professor at the University of Passau, Germany. Since 2015, he has been a professor at Halmstad University, Sweden (now part-time). He received the Ph.D. degree from the Tampere University of Technology, Finland in 2013. He has been the Senior Member of the IEEE since 2012. His areas of interests include vehicular communications and networking, cooperative automated and autonomous driving, future smart mobility solutions.
\end{IEEEbiography}

\end{document}

%% file: acronyms.tex
\begin{acronym}[ACC]
\acro{ACC}{Adaptive Cruise Control}
\end{acronym}
\begin{acronym}[ADS]
\acro{ADS}{automated driving system}
\end{acronym}
\begin{acronym}[AoI]
\acro{AoI}{Age of Information}
\end{acronym}
\begin{acronym}[ARQ]
\acro{ARQ}{Automatic Repeat Request}
\end{acronym}
\begin{acronym}[CACC]
\acro{CACC}{Cooperative Adaptive Cruise Control}
\end{acronym}
\begin{acronym}[CAM]
\acro{CAM}{Cooperative Awareness Message}
\end{acronym}
\begin{acronym}[CBR]
\acro{CBR}{Channel Busy Ratio}
\end{acronym}
\begin{acronym}[CA]
\acro{CA}{Cooperative Awareness}
\end{acronym}
\begin{acronym}[CBF]
\acro{CBF}{Contention-Based Forwarding}
\end{acronym}
\begin{acronym}[CCAM]
\acro{CCAM}{Cooperative, Connected and Automated Mobility}
\end{acronym}
\begin{acronym}[C-ITS]
\acro{C-ITS}{Cooperative Intelligent Transport Systems}
\end{acronym}
\begin{acronym}[CPS]
\acro{CPS}{Collective Perception Service}
\end{acronym}
\begin{acronym}[CPM]
\acro{CPM}{Collective Perception Message}
\end{acronym}
\begin{acronym}[DCC]
\acro{DCC}{Decentralized Congestion Control}
\end{acronym}
\begin{acronym}[DDT]
\acro{DDT}{Dynamic Driving Task}
\end{acronym}
\begin{acronym}[DEN]
\acro{DEN}{Decentralized Environmental Notification}
\end{acronym}
\begin{acronym}[DENM]
\acro{DENM}{Decentralized Environmental Notification Message}
\end{acronym}
\begin{acronym}[DPD]
\acro{DPD}{Duplicate Packet Detection}
\end{acronym}
\begin{acronym}[DPL]
\acro{DPL}{Duplicate Packet List}
\end{acronym}
\begin{acronym}[EDCA]
\acro{EDCA}{Enhanced Distributed Channel Access}
\end{acronym}
\begin{acronym}[EKF]
\acro{EKF}{Extended Kalman Filter}
\end{acronym}
\begin{acronym}[ETSI]
\acro{ETSI}{European Telecommunications Standards Institute}
\end{acronym}
\begin{acronym}[EPAC]
\acro{EPAC}{Electrically Power Assisted Cycle}
\end{acronym}
\begin{acronym}[FCD]
\acro{FCD}{Floating Car Data}
\end{acronym}
\begin{acronym}[ITS]
\acro{ITS}{Intelligent Transport Systems}
\end{acronym}
\begin{acronym}[LocT]
\acro{LocT}{Location Table}
\end{acronym}
\begin{acronym}[LocTE]
\acro{LocTE}{Location Table entry}
\end{acronym}
\begin{acronym}[LSM]
\acro{LSM}{Least Square Method}
\end{acronym}
\begin{acronym}[MC]
\acro{MC}{Maneuver Coordination}
\end{acronym}
\begin{acronym}[MCO]
\acro{MCO}{Multi-channel Operation}
\end{acronym}
\begin{acronym}[MCM]
\acro{MCM}{Maneuver Coordination Message}
\end{acronym}
\begin{acronym}[mmWave]
\acro{mmWave}{millimiter wave}
\end{acronym}
\begin{acronym}[OBU]
\acro{OBU}{On-board Unit}
\end{acronym}
\begin{acronym}[OEDR]
\acro{OEDR}{Object and Event Detection and Reaction}
\end{acronym}
\begin{acronym}[PAM]
\acro{PAM}{Platooning Awareness Message}
\end{acronym}
\begin{acronym}[PCM]
\acro{PCM}{Platooning Control Message}
\end{acronym}
\begin{acronym}[PDR]
\acro{PDR}{Packet-delivery Ratio}
\end{acronym}
\begin{acronym}[QCQP]
\acro{QCQP}{Quadratically Constrained Quadratic Program}
\end{acronym}
\begin{acronym}[RadCom]
\acro{RadCom}{Radar-Based Communication}
\end{acronym}
\begin{acronym}[RSU]
\acro{RSU}{Road-side Unit}
\end{acronym}
\begin{acronym}[RHW]
\acro{RHW}{Road Hazard Warning}
\end{acronym}
\begin{acronym}[SHB]
\acro{SHB}{Single-Hop Broadcasting}
\end{acronym}
\begin{acronym}[S-CBR]
\acro{S-CBR}{Smoothed Channel Busy Ratio}
\end{acronym}
\begin{acronym}[TC]
\acro{TC}{Traffic Class}
\end{acronym}
\begin{acronym}[TDC]
\acro{TDC}{Transmit Data Rate Control}
\end{acronym}
\begin{acronym}[TLS]
\acro{TLS}{Transport Layer Security}
\end{acronym}
\begin{acronym}[TPC]
\acro{TPC}{Transmit Power Control}
\end{acronym}
\begin{acronym}[TRC]
\acro{TRC}{Transmit Rate Control}
\end{acronym}
\begin{acronym}[TTL]
\acro{TTL}{Time-to-Live}
\end{acronym}
\begin{acronym}[VBS]
\acro{VBS}{VRU Awareness Basic Service}
\end{acronym}
\begin{acronym}[VAM]
\acro{VAM}{VRU Awareness Message}
\end{acronym}
\begin{acronym}[VANET]
\acro{VANET}{Vehicular ad hoc Network}
\end{acronym}
\begin{acronym}[VRU]
\acro{VRU}{Vulnerable Road User}
\end{acronym}
\begin{acronym}[V2X]
\acro{V2X}{Vehicle-to-anything}
\end{acronym}
\begin{acronym}[V2V]
\acro{V2V}{Vehicle-to-Vehicle}
\end{acronym}